\documentclass[conference]{IEEEtran}
\IEEEoverridecommandlockouts

\usepackage[table]{xcolor}
\usepackage{graphicx}%
\usepackage{booktabs}
\usepackage{amsmath,amssymb,amsfonts}%
\usepackage{mathrsfs}%
\usepackage{array}
\usepackage{xcolor}%
\usepackage{textcomp}%
\usepackage{manyfoot}%
\usepackage{booktabs}%
\usepackage{algorithm}%
\usepackage{algorithmicx}%
\usepackage{algpseudocode}%
\usepackage{listings}%
\usepackage{textcomp}
\usepackage{tabulary}
\usepackage{tabularx}
\usepackage{hhline}
\usepackage{booktabs} 
\usepackage{arydshln} 
\usepackage{url}
\usepackage{makecell}
\usepackage{hyperref}
\usepackage{multirow}
\usepackage{colortbl} 
\usepackage{color}
\usepackage{longtable} 
\usepackage{pdflscape}
\usepackage{comment}
\usepackage{tablefootnote}
\usepackage{fixfoot}
\usepackage{framed}
\usepackage[many]{tcolorbox}    	
\usepackage{siunitx}
\usepackage{multicol}

\usepackage{balance}

\definecolor{main}{HTML}{5989cf}    
\definecolor{sub}{HTML}{D3D3D3}     

\newtcolorbox[auto counter]{somebox}[1][]{arc=5pt,auto outer arc,left=1pt,boxsep=0.5pt,boxrule=0.5pt,width=\columnwidth,right=1pt, #1}

\usepackage{blindtext}

\begin{document}

\title{Deep Learning and Data Augmentation for Detecting Self-Admitted Technical Debt}
\author{\IEEEauthorblockN{Anonymous Authors}} 




\author{
    \IEEEauthorblockN{
        Edi Sutoyo\IEEEauthorrefmark{1}\IEEEauthorrefmark{2}, Paris Avgeriou\IEEEauthorrefmark{1}, Andrea Capiluppi\IEEEauthorrefmark{1}
    }
    \IEEEauthorblockA{\IEEEauthorrefmark{1}Bernoulli Institute, University of Groningen, Groningen, The Netherlands}
    \IEEEauthorblockA{\IEEEauthorrefmark{2}Department of Information Systems, Telkom University, Bandung, Indonesia}
    Email: \{e.sutoyo, p.avgeriou, a.capiluppi\}@rug.nl
}


\maketitle
\thispagestyle{plain}
\pagestyle{plain}

\begin{abstract} 
Self-Admitted Technical Debt (SATD) refers to circumstances where developers use textual artifacts to explain why the existing implementation is not optimal. Past research in detecting SATD has focused on either identifying SATD (classifying SATD items as SATD or not) or categorizing SATD (labeling instances as SATD that pertain to requirement, design, code, test debt, etc.). However, the performance of these approaches remains suboptimal, particularly for specific types of SATD, such as test and requirement debt, primarily due to extremely imbalanced datasets. To address these challenges, we build on earlier research by utilizing BiLSTM architecture for the binary identification of SATD and BERT architecture for categorizing different types of SATD. Despite their effectiveness, both architectures struggle with imbalanced data. Therefore, we employ a large language model data augmentation strategy to mitigate this issue. Furthermore, we introduce a two-step approach to identify and categorize SATD across various datasets derived from different artifacts. Our contributions include providing a balanced dataset for future SATD researchers and demonstrating that our approach significantly improves SATD identification and categorization performance compared to baseline methods.

\end{abstract}


\maketitle

\section{Introduction}\label{sec1}
Technical debt (TD) is a software development metaphor that represents the extra cost of change as a result of choosing a quick and limited solution over an optimal approach that would, however, take longer to implement \cite{cunningham1992wycash, avgeriou2016managing}. Like financial debt, TD has to be paid back with interest on higher future costs during maintenance and evolution \cite{brown2010managing}. 

In practice, TD is not always visible to all development project stakeholders. In most cases, it is only known to a small number of developers, particularly to those who wrote the code containing the debt. 
To resolve this issue, numerous researchers have proposed various techniques to detect TD and subsequently make it explicit. Most of these prior works have concentrated on detecting TD through static source code analysis \cite{marinescu2004detection, marinescu2010incode, zazworka2014comparing}.

While static code analysis can effectively detect TD in the form of violations of coding rules, dependencies, or smells, another kind of TD is admitted by the developers themselves, known as self-admitted technical debt (SATD). SATD refers to circumstances where developers use code comments, issues, pull requests, or other textual artifacts to explain why the existing implementation is not optimal~\cite{potdar2014exploratory}. For example, developers frequently use terms, e.g., `\textit{TODO}' or `\textit{fixme}', when admitting the existence of TD in source code comments. The key benefit of SATD is that, unlike approaches utilizing static code analysis to detect \textit{proxies} of TD (such as dependencies or smells), it represents TD \textit{per se} as stated by developers who are familiar with the code \cite{huang2018identifying}. Furthermore, the approaches grounded in source code analysis encounter significant challenges with false positives \cite{fontana2016antipattern}, wherein numerous source code components are identified as problematic despite not being perceived as such by developers. This is primarily because these methods solely depend on the structure of the source code for detecting code smells, disregarding crucial factors such as developer feedback, project domain, and contextual information surrounding the detection of code smells \cite{da2017using}.

The detection of SATD in the literature has, in general, pursued two main directions: 
\begin{enumerate}
    \item A binary classification (i.e., \textit{identification}) of a software development artifact, into either `SATD' or `Not-SATD'. 
    This is the simplest classification, but it still lists all SATD items and allows the development team to understand how much SATD exists in the system \cite{dai2017detecting}.
    \item A multi-type classification (i.e., \textit{categorization}) of artifacts using labels related to the software development activities (e.g., `requirement debt', `design debt', `code debt') \cite{wattanakriengkrai2019automatic, xavier2020beyond, zhu2023scgru}. Recognizing the development activities where SATD occurs has a long-term advantage: 
    each specific type of technical debt requires different repayment strategies \cite{alves2014towards}, thus categorizing the specific types of SATD, and monitoring their evolution, facilitates their repayment. Furthermore, identifying specific types of SATD is essential for developers to understand the distribution of the different types better \cite{maldonado2015detecting} as well as the effects of the different types on maintenance efficiency \cite{chen2021multiclass}.

\end{enumerate}

The most recent attempts at detecting SATD have included various software artifacts to label different types of SATD effectively~\cite{dai2017detecting, li2022identifying, xavier2022documentation}, this has translated into a multi-dimensional problem, where specific artifacts are starting to be used in the detection of specific types of SATD. Recently, Li et al. \cite{li2023automatic} proposed an automatic SATD detection approach across multiple sources: source code comments, issue trackers, commit messages, and pull requests. However, the performance of such approach (i.e., F1-score) needs to be improved to categorize SATD more accurately; for example, the approach of Li et al. \cite{li2023automatic} struggles in effectively categorizing certain SATD types, e.g., test and requirement debt. 


The performance issue arises from a common challenge encountered in classification tasks: the datasets used for training SATD classification are imbalanced since there is a significant disparity in the number of instances belonging to different classes \cite{provost2013data}. The imbalance occurs due to the set of labels resulting from manual data annotation, where one class significantly outweighs the others. This skewed distribution can make it difficult for the model to effectively learn the minority class \cite{chen2021multiclass}. 

This paper attempts (a) to contribute the steps to rectify imbalanced SATD datasets and (b) to achieve better performance at both \textit{identifying} and \textit{categorizing} SATD. To achieve (a), we utilize an automated approach that can effectively increase the variety of instances, improve the representation of features, and balance the available datasets. We demonstrate the significance of the proposed approach in terms of F1-score and compare it to baseline models. To achieve (b), we present a two-step approach to identify and categorize SATD from multiple software artifacts: in the first step, a deep learning architecture called BiLSTM is utilized to \textit{identify} SATD and to separate it from Not-SATD. Using such identified SATD, the second step utilizes the BERT architecture to \textit{categorize} different types of SATD. 

The main contributions of this research are the following:
\begin{enumerate}

    \item We employ a novel approach for \textit{identifying} and \textit{categorizing} SATD from four different artifacts in software systems, utilizing a combination of deep learning architectures and data augmentation techniques. This approach contributes to the software engineering field by providing a new method for detecting and managing SATD in software systems. 

    \item To address the issue of \textit{data imbalance}, we utilize a large language model (LLM) to effectively increase the variety of instances and improve feature representation. To the best of our knowledge, this is the first work that employs data augmentation to overcome imbalanced classes in SATD identification and categorization. We make this balanced dataset available for further studies \cite{satdaug}.

    \item We conduct comprehensive experiments to evaluate the performance of our proposed approach and compare it to baselines. The results indicate that our approach outperforms the baselines.

\end{enumerate}

This study is organized as follows: Section~\ref{sec2} introduces the background, while Section~\ref{sec3} describes the study design. Section~\ref{sec4} reports and elaborates on the obtained results, while Section~\ref{sec5} discusses our results. Finally, Section~\ref{sec6} comprehensively summarizes the findings and future work.


\section{Background}\label{sec2}
Previous studies have demonstrated that technical debt (TD), while impairing future maintenance and evolution, is ubiquitous and unavoidable in software development processes \cite{lim2012balancing, wehaibi2016examining}. Developers must spend additional time and effort building new features or fixing bugs to either pay off the debt (e.g., through refactoring) or pay the interest on it.


Researchers have focused on detecting TD for over a decade, as detection is the first step in managing it \cite{li2015systematic}. Various approaches have been proposed since TD can lower software quality, pose long-term risks, and require to perform redesign and refactoring. Unlike TD detection methods which only focus on the source code, SATD detection has many advantages because SATD items are directly and intentionally recognized by developers, e.g., through code comments \cite{da2017using}. Furthermore, SATD represents the debt \textit{per se} as it was introduced by the developer, in contrast to debt identified by static source code analysis, which is only a proxy \cite{sala2021debthunter}. 

For almost a decade, SATD detection from source code comments has rapidly evolved, whether it is the identification of SATD and Not-SATD (binary class) \cite{da2017using, guo2021far, sridharan2021data} or categorization of specific types of SATD (multi-class) \cite{zhu2023scgru, chen2021multiclass, zhu2021detecting, yu2022exploiting}. Furthermore, researchers have also been successful in identifying the existence of SATD from other artifacts, namely issue tracking systems, commit messages, and pull requests \cite{dai2017detecting, xavier2020beyond, li2022identifying, zampetti2021self}. 

Recently, Li et al. \cite{li2023automatic} proposed an approach to categorize specific types of SATD (i.e., code/design, requirement, documentation, and test debt) by mining four different artifacts, namely source code comments (CC), issues section (IS), pull section (PS), and commit messages (CM). Their approach achieved an average F1-score of 0.611.  Furthermore, they have stated that their approach still struggles to categorize test and requirement debt. In our work, we attempt to enhance performance by optimizing the categorization of both test and requirement debt, as well as the other SATD types. 

In a recent literature review~\cite{sutoyo2023detecting}, 68 papers were analyzed concerning the detection of SATD. It was found that BiLSTM and BERT were the two deep learning architectures that achieved the highest F1-score values for identifying and categorizing SATD, respectively. Despite previous research indicating the superior performance of both architectures, they were found to struggle with imbalanced data. Therefore, in this study, we propose a two-step approach by leveraging the capabilities of BiLSTM and BERT, integrated with AugGPT data augmentation, to effectively address the challenge posed by imbalanced data for SATD identification and categorization.

The datasets used to produce those performance comparison results were mostly sourced from two main references: Maldonado et al. \cite{da2017using}, and Guo et al. \cite{guo2021far}. Their code comments datasets are summarized as follows: the first is a multi-class SATD dataset containing five specific types of SATD\footnote{\url{https://github.com/maldonado/tse.satd.data}} (`design debt', `requirement debt', `defect debt', `documentation debt', and `test debt') plus the `Not-SATD' label, where instances were labeled as not containing any SATD. The imbalance of these labels is noticeable: for example, `design debt' contains 2,703 instances, but the `documentation debt' is only 54. The second dataset~\cite{guo2021far} contains two labels (SATD and Not-SATD), and it has been reused by other researchers for the identification (but not categorization) of SATD\footnote{\url{https://github.com/Naplues/MAT}}.

\section{Study Design}\label{sec3}
\begin{figure}[] 
\centerline{\includegraphics[width=0.33\textwidth]{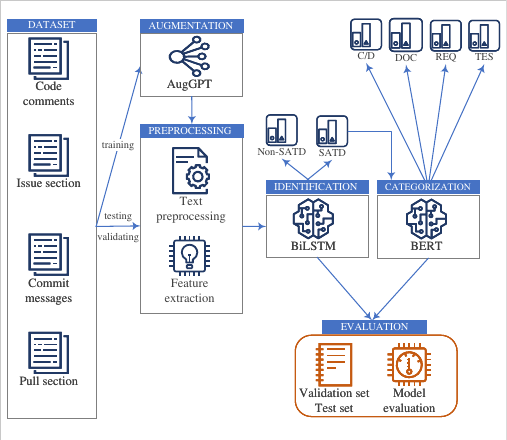}}
\caption{Research Methodology}
\label{fig_methodology}
\end{figure}

The objectives of this study are to identify SATD and categorize its types when considering various artifacts.
The methodology shown in Figure~\ref{fig_methodology} contains 6 parts: 
\begin{itemize}
    \item (Part i - Dataset) -- We consider four artifacts to detect SATD: source code comments (CC), issues section (IS), commit messages (CM), and pull section (PS). Our approach can be extended to any data source that might potentially contain traces of SATD. See Subsection~\ref{part-a}.
    \item (Part ii - Data Augmentation) -- We employ an LLM data augmentation strategy to address data imbalances. The specific label instances will be augmented to balance their examples and avoid skewness in the representation of the categories. Data augmentation is applied exclusively to the training set to prevent data leakage. See Subsection~\ref{part-b}.
    \item (Part iii - Preprocessing) -- We treat all software artifacts with text preprocessing and feature extraction. Text preprocessing is used to eliminate noise in the data. Meanwhile, feature extraction is utilized to enable an algorithm to learn from training data with pre-defined features before it is fed into a deep learning architecture. See Subsection~\ref{part-c}, and \ref{part-d}.
    \item (Part iv - Identification) -- The first of our two-step approach is a binary classification of SATD: we identify whether the four sources of artifacts contain either SATD or Not-SATD instances. See Subsection~\ref{part-e}.
    \item (Part v - Categorization) -- The second step in our two-step approach only considers the instances that contain SATD, as identified in the previous step: by utilizing BERT, we categorize SATD types, namely code/design debt (\textbf{C/D}), documentation debt (\textbf{DOC}), test debt (\textbf{TES}), and requirement debt (\textbf{REQ}). See Subsection~\ref{part-f}.
    \item (Part vi - Evaluation) -- To evaluate the effectiveness of the proposed approach, we run experiments on a publicly available dataset provided by Li et al. \cite{li2023automatic}. The F1-score is employed to evaluate the performance of the proposed approach. We employ a stratified train-validation-test split for each of the artifacts, allocating 80\% of the data for training, 10\% for validation, and 10\% for testing. See Subsection~\ref{part-g}. 
\end{itemize}  



\subsection{Research Questions}
Our work is based on three research questions (RQs):

\textbf{RQ1}: \textit{How does the proposed approach impact the overall performance in the  identification and categorization of SATD?}\\
Motivation: this RQ aims to determine whether the two-step approach can enhance the performance of baseline approaches from the literature.

\textbf{RQ2}: \textit{How does augmenting datasets to address highly imbalanced data impact the overall performance?}\\ 
Motivation: the purpose of this RQ is to investigate the effect that data augmentation techniques have on the performance of our approach. 

\textbf{RQ3}: \textit{What are the most indicative keywords that represent each source of artifacts and specific types of SATD?}\\
Motivation: detecting and categorizing SATD through highly specific keywords can be enhanced to lead to a more comprehensive understanding of SATD and its impact on software development. 
Since existing approaches have limitations in identifying SATD (i.e., low F1-scores), researchers can categorize SATD patterns and trends using keywords across different artifacts. In addition, the identified keywords can be used to improve the explainability of our model's prediction results.





To answer these RQs, we follow the approach presented in Fig.~\ref{fig_methodology}. Below, we will provide a more comprehensive elaboration on each of its steps.

\subsection{Training and testing datasets}
\label{part-a}

The datasets from Maldonado et al.~\cite{da2017using} and Guo et al.~\cite{guo2021far} indicate that instances can simply be identified as SATD and Not-SATD but can also be categorized into specific types of SATD. These two research datasets have been used so far a total of 34 times by other researchers to identify SATD and to evaluate machine learning algorithms, as reported in \cite{sutoyo2023detecting}. However, these datasets focus on SATD detection by only using source code comments. 


To evaluate our approach, we utilize the dataset provided by Li et al. \cite{li2023automatic}, derived from various artifacts (see Table~\ref{tab:tb_dataset}). This dataset includes 5,000 commit messages and 5,000 pull requests from 103 Apache open-source projects, as well as 4,200 issues from 7 open-source projects using two issue tracking systems (Jira and Google Monorail). Each instance was manually annotated to identify whether it was Not-SATD or SATD, and further categorized into specific types of SATD. Additionally, they analyzed 23.6M source code comments, 1.3M commit messages, 3.7M issue sections (including individual issue summaries, descriptions, or comments), and 1.7 million pull request sections (including summaries, descriptions, or comments) from 103 open-source projects to categorize SATD. As shown in Table~\ref{tab:tb_dataset}, the CC artifact contains the same data as in \cite{da2017using}\footnote{\url{https://github.com/yikun-li/satd-different-sources-data}}.


To apply the two-step approach to this dataset, first, we merge all types of SATD (C/D, DOC, TES, and REQ) into one class, namely “SATD”, so that there are only two classes (SATD and Not-SATD). This aims to conduct an initial \textit{identification} step of SATD with the BiLSTM. If the item is identified as SATD, the second step then uses the BERT to \textit{categorize} it into the specific types of SATD, as shown in the rows of Table~\ref{tab:tb_dataset}.



\begin{table}[ht]
    \caption{Dataset from Li et al. \cite{li2023automatic}} 
    \begin{tabular}{p{1.7cm} p{1.2cm} p{1.2cm} p{1.2cm} p{1.2cm}}
        \toprule
        SATD Type & CC & IS & PS & CM\\
        \midrule
        
        C/D & 2,703 & 2,169 & 510 & 522\\
        DOC & 54 & 487 & 101 & 98\\
        TES & 85 & 338 & 68 & 58\\
        REQ & 757 & 97 & 20 & 27\\\hline
        SATD & 3,599 & 3,091 & 699 & 705\\
        Not-SATD & \textbf{58,204} & \textbf{20,089 }& \textbf{4,301} & \textbf{4,295}\\
        \bottomrule
    \end{tabular}
    \label{tab:tb_dataset}
\end{table}

\subsection{Data Augmentation} 
\label{part-b}

\label{sec:_data_augmentation}
Based on our examination of the dataset in Table~\ref{tab:tb_dataset}, we observed significant imbalances in the distribution of specific SATD types. For instance, in the dataset derived from CC artifacts, DOC debt comprises only 54 rows, accounting for 1.99\% compared to C/D debt and 0.09\% compared to Not-SATD. Similar imbalances exist in datasets from issue trackers, pull requests, and commit messages.

The situation of extremely imbalanced data is a typical challenge that needs to be addressed when detecting SATD, particularly for categorizing the specific SATD types \cite{sridharan2021data}. 
The issue of class imbalance can significantly affect how SATD instances are classified \cite{chen2021multiclass}. Furthermore, it is challenging for deep learning architectures to categorize correct SATD types since they learn semantic information from a limited number of data points. As a result, dealing with the class imbalance and increasing the number of data in the minor class is necessary.

Undersampling and oversampling are common approaches to resolving imbalance issues \cite{drummond2003c4}, but both have shortcomings \cite{yap2014application}. Undersampling reduces samples from large classes to match smaller classes, potentially deleting crucial information and limiting sample variety while decreasing training samples. Oversampling balances data by repeatedly selecting samples from small classes, which does not enhance sample variety and often leads to overfitting \cite{lee2018oversampling}.

Data augmentation, another key method to address imbalance, involves generating new samples based on existing ones. This not only increases the number of samples in small classes but also enhances sample diversity. Traditional techniques include synonym replacement and random operations \cite{feng2021survey}, while recent methods like back-translation and word vector interpolation have been explored \cite{sennrich2015improving, jindal2020augmenting}. However, these approaches still struggle with maintaining accuracy and diversity and often require human annotation \cite{feng2021survey}. Therefore, in this study, we utilized AugGPT \cite{dai2023chataug}, based on ChatGPT-3.5, as a powerful data augmentation strategy to address the issue of imbalanced classes. AugGPT demonstrates high faithfulness and compactness in preserving accuracy compared to other augmentation methods \cite{dai2023chataug}. This technique generates text by probabilistic sampling from a language's lexicon \cite{brown2020language}, effectively enhancing dataset diversity and addressing the challenges posed by imbalanced classes in SATD detection. 






In this paper, AugGPT was employed to produce a supplementary training dataset with the primary objective of generating paraphrased text for each data point while ensuring the original meaning remained unchanged. Preserving the meaning is crucial to avoid potential mislabeling. According to Dai et al. \cite{dai2023chataug}, single-turn and multi-turn dialogues are two design prompts for the data augmentation process. This study formulated the prompt instruction through experimentation and trial-and-error, guided by the principles outlined in the multi-turn prompt dialogue as proposed by them \cite{dai2023chataug}.

\begin{figure}[ht]
\centerline{\includegraphics[width=0.33\textwidth]{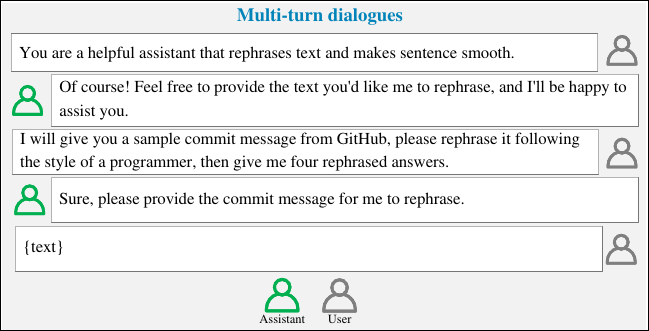}}
\caption{Multi-turn dialogues prompt for CM artifact}
\label{fig_dialogue}
\end{figure}

As depicted in Fig.~\ref{fig_dialogue}, we also added context (e.g., specifying “commit message from GitHub”) and a persona (i.e., a programmer) to ensure the generated augmentation data closely resembled the original text \cite{white2023prompt}. This approach effectively augmented the dataset, enhancing its diversity and introducing linguistic variation.

To illustrate the augmentation process, let's examine an example from the dataset derived from the CM artifact, as outlined in Table~\ref{tab:tb_augmentation_process}. Each specific type of SATD received varying degrees of data augmentation to balance the total number with the largest SATD type (i.e., C/D debt with 522 instances). For instance, DOC debt, initially comprising 98 instances, underwent fourfold augmentation to align with the amount of C/D debt. The AugGPT multi-turn prompt dialogue method was employed to achieve this augmentation. Similarly, TES debt was augmented eightfold, resulting in 522 instances, and REQ debt underwent eighteenfold augmentation, yielding 513 instances. C/D debt and Not-SATD categories were not subjected to data augmentation.

\begin{table}[htpb]
    \caption{Dataset from CM artifact} 
    \begin{tabular}{p{1.5cm} p{1.5cm} p{2.6cm} p{1.3cm}}
        \toprule
        \multirow{2}{6em}{SATD Type} & Original amount & \# of times each item is augmented & Final amount\\
        \midrule
        C/D & 522 & - & 522\\
        DOC & 98 & 4 & 490\\
        TES & 58 & 8 & 522\\
        REQ & 27 & 18 & 513\\

        \bottomrule
    \end{tabular}
    \label{tab:tb_augmentation_process}
\end{table}

As a result of the augmentation of all the artifacts, the dataset generated through data augmentation utilizing AugGPT used in this study is available in the replication package \cite{satdaug}, and its metadata is shown in Table~\ref{tab:tb_augmented_dataset} below. 

\begin{table}[htpb]
    \caption{Original and \colorbox{blue!20}{augmented} for each SATD type based on Table~\ref{tab:tb_dataset}} 
    \begin{tabular}{p{1.8cm} p{1.2cm} p{1.2cm} p{1.2cm} p{1.2cm}}
        \toprule

        SATD Type & CC & IS & PS & CM\\
        \midrule
        C/D & 2,703 & 2,169 & 510 & 522\\
        DOC & \cellcolor{blue!25}2,701 & \cellcolor{blue!25}1,948 & \cellcolor{blue!25}505 & \cellcolor{blue!25}490\\
        TES & \cellcolor{blue!25}2,635 & \cellcolor{blue!25}2,028 & \cellcolor{blue!25}476 & \cellcolor{blue!25}522\\
        REQ & \cellcolor{blue!25}2,271 & \cellcolor{blue!25}2,134 & \cellcolor{blue!25}500 & \cellcolor{blue!25}513\\

        \bottomrule
    \end{tabular}
    \label{tab:tb_augmented_dataset}
\end{table}

To see the significance of the augmentation process, Table~\ref{tab:tb_shannon} shows the outcomes of applying the Shannon entropy \cite{shannon2001mathematical} to evaluate the class distribution pre- and post-augmentation across the datasets. The table demonstrates an enhancement in the balance of the datasets concerning both SATD Identification and Categorization.

\begin{table}[ht]
    \caption{Class distribution calculation using Shannon Entropy \cite{shannon2001mathematical}} 
    \begin{tabular}{p{1cm} p{1.4cm}|p{1.4cm} p{1.4cm}|p{1.4cm}}
        \toprule
        \multirow{3}{4em}{Artifact} & \multicolumn{2}{c}{\textbf{Identification}} & \multicolumn{2}{c}{\textbf{Categorization}}\\
        \cmidrule{2-5}
        & Original & Augmented & Original & Augmented\\
        \midrule
        CC & 0.231 & 0.611 & 0.500 & 0.998\\
        IS & 0.569 & 0.873 & 0.642 & 0.999\\
        PS & 0.585 & 0.901 & 0.604 & 0.999\\
        CM & 0.589 & 0.909 & 0.596 & 0.999\\
        \bottomrule
    \end{tabular}
    \label{tab:tb_shannon}
\end{table}

\subsection{Text Preprocessing}
\label{part-c}\
Developers often use their own style and preferences when adding comments to source code, issue trackers, commit messages, and pull requests, resulting in diverse text formats. These texts provide valuable information, such as reasons for code changes, progress updates on group efforts, or documentation of modifications \cite{de2006documentation}. Since machine learning and deep learning algorithms require structured text, preprocessing is essential to reduce data noise.

This study employed standard preprocessing procedures, including data cleansing, duplicate removal, lowercase conversion, tokenization, stop word removal, punctuation removal, and lemmatization. Additional steps included excluding short words (two letters or fewer), removing numbers, URLs, and non-ASCII characters, and eliminating extra white spaces.

\subsection{Feature Extraction}
\label{part-d}
Feature extraction using word embeddings refers to the process of utilizing pre-trained word embeddings to represent words in a text or document as numerical features that can be used for various machine learning tasks \cite{chowdhary2020natural}. 


Instead of training word embeddings from scratch, pre-trained embeddings like GloVe \cite{pennington2014glove} and BERT \cite{devlin2018bert} are often used to extract meaningful word representations. These embeddings predict a word's context based on its neighboring words or reconstruct the word from its context, encoding semantic information and enabling algorithms to understand and infer language. Therefore, this study uses GloVe embeddings for BiLSTM and BERT embeddings for BERT architectures.

After data augmentation, text preprocessing, and feature extraction, we employ two architectures: BiLSTM (the \textit{Identification} in Fig.~\ref{fig_methodology}) is used to identify SATD or Not-SATD artifacts (binary classification), while BERT (the \textit{Categorization}) is utilized to categorize specific types of SATD (multi-class classification). After these steps, we will evaluate and compare how well each architecture performs.

\subsection{SATD Identification with BiLSTM}
\label{part-e}


According to the findings from a prior study performed to identify the optimal algorithms to identify SATD~\cite{sutoyo2023detecting}, BiLSTM \cite{schuster1997bidirectional} is the most effective deep learning architecture and has been routinely utilized by researchers for this purpose. Leveraging the results of past research as the first step of our proposed approach, we utilize the BiLSTM architecture to identify whether an item from a specific artifact should be considered as either SATD or Not-SATD. 

The model's initial layer is an embedding layer, configured with a vocabulary size, an embedding dimension, and a pre-trained embedding matrix. The core architecture consists of multiple stacked BiLSTM layers. The first BiLSTM layer has 128 units and is accompanied by a Dropout layer with a rate of 0.3 to reduce overfitting. A Batch Normalization layer follows to stabilize and accelerate training by normalizing activations. Next, two additional BiLSTM layers are stacked, each with a Dropout layer at a 0.3 rate. The second BiLSTM layer has 64 units, while the third returns to 128 units. The final Bidirectional LSTM layer, with 128 units, consolidates the learned features. The training was monitored for validation loss and halted early when the minimum validation loss was achieved.

\subsection{SATD Categorization with BERT}
\label{part-f}
A pre-trained language model, specifically the BERT-base-uncased implemented in the Transformers library\footnote{\url{https://huggingface.co/bert-base-uncased}}, was utilized for the purpose of categorizing SATD. BERT-base was selected based on its established superiority in categorizing SATD compared to other machine learning and deep learning models \cite{sutoyo2023detecting}. Additionally, the BERT-base demonstrates comparatively lower computational resource requirements in terms of time and memory when compared to the BERT-Large model. 




The BERT-base-uncased model was pre-trained using a vast collection of English text with a self-supervised approach. The classifier consists of a sequential combination of layers: a linear layer mapping the BERT output to the hidden layer, a ReLU activation function to introduce non-linearity, and another linear layer mapping the hidden layer to the output layer. To optimize the pre-trained BERT-base-uncased model (768 hidden units, 12 layers, and 110M parameters) for categorizing SATD, we used CrossEntropyLoss as the loss function for the multi-class classification task. The model was trained using the AdamW optimizer, a modified version of the Adam optimizer that incorporates weight decay \cite{loshchilov2017decoupled}. 


To further optimize our model's performance, we fine-tuned several hyperparameters\footnote{We experimented with different values for each parameter, and the results are available in the replication package.}. Specifically, we used a batch size of 32, determining the number of samples processed in a single iteration. The ReLU activation function was employed to enhance the model's learning and prediction capabilities. We set the learning rate to 5e-5, guiding the model on how quickly to adjust its internal parameters during training. Additionally, we selected an AdamW optimizer with an epsilon value of 1e-8 to ensure stability and efficiency in the optimization process. These fine-tunings collectively enhanced our model's performance.

Experiments were conducted on all datasets, including CC, IS, PS, and CM artifacts. The deep learning algorithms were implemented using the PyTorch library and trained on NVIDIA Tesla V100 GPUs.

\subsection{Model Evaluation}
\label{part-g}
We use the F1-score ($\frac{2*Precision*Recall}{Precision+Recall}$) to evaluate the performance of our approach because it is computed as the harmonic mean of precision and recall\footnote{Other metric alternatives, like the standard AUC-ROC curve or MCC, are only for binary classification tasks and can only be applied for identification, not categorization.}. This makes it more sensitive to extreme values than other metrics like the arithmetic mean, which can be helpful when dealing with imbalanced datasets. 




\section{Results}\label{sec4}
This section provides a detailed analysis of the outcomes from our proposed two-step approach. For RQ1, we utilize the BiLSTM and BERT architectures without data augmentation for SATD identification and categorization. To address RQ2, we apply the BiLSTM and BERT architectures with data augmentation techniques for these tasks. The classification focuses on the four SATD types presented by \cite{li2023automatic}. Finally, we elaborate on the most representative SATD keywords to address RQ3.



\subsection{RQ1: SATD identification and categorization without data augmentation (base)}
\label{step-1}
To demonstrate the significance of our approach in identifying SATD and in categorizing various types of SATD as measured by F1-score, in this subsection, we will first discuss the results of BiLSTM and BERT without data augmentation.

To our knowledge, no previous studies have automated the identification of SATD (binary classification) on various artifacts. To establish baselines for comparison, we used studies that employed the datasets from \cite{da2017using, guo2021far}, which solely examine CC artifacts. We utilized and modified the source code provided by \cite{chen2021multiclass, guo2021far, sridharan2021data, yin2023two} to enable its execution for identification and categorization across IS, PS, and CM artifacts. These modifications included adjusting the class numbers, standardizing the preprocessing steps, and standardizing the split ratio for training and testing. Therefore, to showcase the effectiveness of our approach, we compared it against several existing approaches: Natural Language Processing (NLP) \cite{da2017using}, the pioneering SATD detection; and Matches task Annotation Tags (MAT) \cite{guo2021far}, a straightforward approach primarily applied to CC artifact. Additionally, we evaluated the performance of eXtreme Gradient Boosting+Synthetic Minority Oversampling Technique (XGBoost+SMOTE) \cite{sridharan2021data}, which incorporates oversampling techniques; eXtreme Gradient Boosting+Easy Data Augmentation (XGBoost+EDA) \cite{chen2021multiclass}, which integrates data augmentation strategies; LightGBM \cite{yin2023two}, which has been optimized for handling sparse data; JSD-GAN \cite{yu2023detecting}, which employs deep learning data augmentation method, namely Generative Adversarial Network (GAN); and BiLSTM for SATD identification across all artifacts.

In terms of SATD categorization, our study compared the proposed approach with XGBoost+SMOTE \cite{sridharan2021data}, XGBoost+EDA \cite{chen2021multiclass}, JSD-GAN \cite{yu2023detecting}, MT-Text-CNN \cite{li2023automatic}, and BERT. XGBoost+SMOTE and XGBoost+EDA were chosen as baselines because both use data augmentation approaches and are capable of performing categorization tasks. Meanwhile, the JSD-GAN model represents the latest approach for SATD categorization. 



In the identification task (see Table~\ref{tab:tb_bilstm_result}), concerning CC artifacts, BiLSTM (base) achieves the second position relative to JSD-GAN, with respective scores of 0.875 and 0.910. Additionally, BiLSTM (baseline) exhibits superior performance compared to all other artifacts by attaining Macro average F1-scores of 0.748, 0.668, and 0.797 for IS, PS, and CM artifacts, respectively. These findings align with F1-score outcomes reported in studies utilizing datasets provided by \cite{da2017using} and \cite{guo2021far}.



Regarding SATD categorization, the results presented in Table~\ref{tab:tb_bert_result} show that BERT outperformed XGBoost+SMOTE, XGBoost+EDA, JSD-GAN, and MT-Text-CNN in categorizing C/D, DOC, and TES debts (excluding commit messages). However, BERT did not achieve better results for REQ debt categorization. In terms of the Macro-averaged F1-score, BERT surpassed the baseline methods, except in the CM artifact. Specifically, BERT achieved Macro-averaged F1-score values of 0.656 (compared to 0.619 with MT-Text-CNN) for CC artifacts, 0.719 (0.453) for IS artifacts, 0.455 (0.441) for PS artifacts, and 0.448 (0.475) for CM artifacts.

Overall, based on the findings presented in Tables~\ref{tab:tb_bilstm_result} and ~\ref{tab:tb_bert_result}, it appears that BiLSTM architecture demonstrates only a slight enhancement in the identification task, whereas BERT architecture exhibits notably superior performance, albeit not universally across all SATD types in terms of F1-score results. 

\begin{table}[htpb]
    \caption{Results of SATD Identification-(\colorbox{gray!25}{These} scores show BiLSTM improves the baselines. 
    \colorbox{blue!25}{These} scores show BiLSTM+AugGPT improve the baselines)} 
    \begin{tabular}{p{0.5cm} p{2.7cm} p{1.22cm} p{0.9cm} p{1.37cm}}
        \toprule
        \multirow{3}{4em}{Artifact   } & \multirow{3}{11em}{ Experiments} & \multicolumn{3}{c}{F1-score}\\
        \cmidrule{3-5}
         & &Not-SATD&SATD&Macro-Avg.\\
        \midrule
        \multirow{8}{4em}{CC} 
        & NLP \cite{da2017using} & 0.929 & 0.494 & 0.711\\ 
        & MAT \cite{guo2021far} & 0.985 & 0.721 & 0.853\\ 
        & XGBoost+SMOTE \cite{sridharan2021data} & 0.980 & 0.540 & 0.760\\
        & XGBoost+EDA \cite{chen2021multiclass} & 0.989 & 0.721 & 0.855\\
        & LightGBM \cite{yin2023two} & 0.927 & 0.728 & 0.828\\
        & JSD-GAN \cite{yu2023detecting} & 0.989 & 0.831 & 0.910 \\
        & BiLSTM (base) & 0.952 & 0.799 & 0.875\\
        & \textbf{BiLSTM+AugGPT} & 0.952 & \cellcolor{blue!25}0.927 & \cellcolor{blue!25}0.939\\
        
        \midrule
        \multirow{6}{4em}{IS}
        & XGBoost+SMOTE \cite{sridharan2021data} & 0.893 & 0.475 & 0.684\\
        & XGBoost+EDA \cite{chen2021multiclass} & 0.919 & 0.453 & 0.686\\
        & LightGBM \cite{yin2023two} & 0.946 & 0.492 & 0.719\\
        & JSD-GAN \cite{yu2023detecting} & 0.924 & 0.390 & 0.657 \\
        & BiLSTM (base) & 0.937 & \cellcolor{gray!25}0.559 & \cellcolor{gray!25}0.748\\
        & \textbf{BiLSTM+AugGPT} & 0.937 & \cellcolor{blue!25}0.820 & \cellcolor{blue!25}0.878\\

        \midrule
        \multirow{6}{4em}{PS} 
        & XGBoost+SMOTE \cite{sridharan2021data} & 0.866 & 0.381 & 0.623\\
        & XGBoost+EDA \cite{chen2021multiclass} & 0.878 & 0.373 & 0.625\\
        & LightGBM \cite{yin2023two} & 0.938 & 0.429 & 0.683\\
        & JSD-GAN \cite{yu2023detecting} & 0.927 & 0.331 & 0.629 \\
        & BiLSTM (base) & 0.915 & 0.422 & 0.668\\
        & \textbf{BiLSTM+AugGPT} & 0.917 & \cellcolor{blue!25}0.806 & \cellcolor{blue!25}0.862\\
        
        \midrule
        \multirow{6}{4em}{CM} 
        & XGBoost+SMOTE \cite{sridharan2021data} & 0.937 & 0.642 & 0.790\\
        & XGBoost+EDA \cite{chen2021multiclass} & 0.937 & 0.429 & 0.683\\
        & LightGBM \cite{yin2023two} & 0.934 & 0.416 & 0.675\\
        & JSD-GAN \cite{yu2023detecting} & 0.915 & 0.191 & 0.553 \\
        & BiLSTM (base) & 0.926 & \cellcolor{gray!25}0.668 & \cellcolor{gray!25}0.797\\
        & \textbf{BiLSTM+AugGPT} & \cellcolor{blue!25}0.940 & \cellcolor{blue!25}0.821 & \cellcolor{blue!25}0.880\\

        \bottomrule
    \end{tabular}
    \label{tab:tb_bilstm_result}
\end{table}






\begin{somebox}
  \textbf{Answer to RQ1:}  \\
  BiLSTM (base) generally outperforms the literature baselines in SATD detection, especially for the SATD class. The BERT model only partially improves SATD categorization and not for all SATD types.
\end{somebox}

\subsection{RQ2: SATD identification and categorization with AugGPT}
\label{step-2}
In the identification task, BiLSTM+AugGPT demonstrates superior performance compared to standard BiLSTM in terms of F1-score. As illustrated in Table~\ref{tab:tb_bilstm_result}, BiLSTM+AugGPT achieves Macro-averaged F1-score values of 0.939 (compared to 0.875 with regular BiLSTM) for CC artifacts, 0.878 (0.748) for IS artifacts, 0.862 (0.668) for PS artifacts, and 0.880 (0.797) for CM artifacts. The Macro-averaged F1-score shows that BiLSTM+AugGPT not only improves SATD detection but also maintains a balanced performance across both classes.

The second step of our approach uses the BERT+AugGPT, so we compared the F1-score of the MT-Text-CNN \cite{li2023automatic} and other aforementioned baselines. This second step aims to properly categorize specific types of SATD (i.e., C/D, DOC, TES, and REQ). 

As demonstrated in Table~\ref{tab:tb_bert_result}, based on Macro-averaged F1-score, the overall findings suggest that BERT+AugGPT exhibits a notably stronger performance in comparison to MT-Text-CNN~\cite{li2023automatic} for SATD categorization across all sources of artifacts. More specifically, the BERT+AugGPT outperforms the XGBoost+SMOTE \cite{sridharan2021data}, XGBoost+EDA \cite{chen2021multiclass}, JSD-GAN \cite{yu2023detecting}, MT-Text-CNN \cite{li2023automatic}, and BERT in categorizing all the specific types of SATD in all datasets. It maintained high F1-scores consistently, with the highest Macro Avgs. for CC (0.882), IS (0.899), PS (0.876), and CM (0.847).





\begin{table}[htpb]
    \caption{Results of SATD Categorization-(\colorbox{gray!25}{These} scores show BERT improves the baselines. \colorbox{blue!25}{These} scores show BERT+AugGPT improve the baselines)} 
    \begin{tabular}{p{1cm} p{0.9cm} p{0.9cm} p{0.9cm} p{0.9cm} p{1.5cm}}
        \toprule
        \multirow{3}{7em}{Artifact} & \multicolumn{5}{c}{F1-score}\\
        \cmidrule{2-6}
         & C/D & DOC & TES & REQ & Macro-Avg.\\
        \midrule
        \multicolumn{6}{c}{XGBoost+SMOTE \cite{sridharan2021data}}\\
        \midrule
        CC & 0.878 & 0.571 & 0.571 & 0.333 & 0.589\\
        IS & 0.889 & 0.706 & 0.667 & 0.308 & 0.642\\
        PS & 0.863 & 0.571 & 0.364 & 0.000\footnotemark{} & 0.450\\
        CM & 0.886 & 0.800 & 0.444 & 0.500 & 0.658\\

        \midrule
        \multicolumn{6}{c}{XGBoost+EDA \cite{chen2021multiclass}}\\
        \midrule
        CC & 0.884 & 0.667 & 0.269 & 0.500 & 0.580\\ 
        IS & 0.887 & 0.637 & 0.761 & 0.462 & 0.687\\ 
        PS & 0.754 & 0.472 & 0.545 & 0.211 & 0.495\\ 
        CM & 0.667 & 0.533 & 0.516 & 0.286 & 0.500\\ 
        \midrule

        \multicolumn{6}{c}{JSD-GAN \cite{yu2023detecting}}\\
        \midrule
        CC & 0.904 & 0.667 & 0.200 & 0.590 & 0.590 \\
        IS & 0.814 & 0.577 & 0.545 & 0.111 & 0.512\\
        PS & 0.825 & 0.400 & 0.636 & 0.000\footnotemark[\value{footnote}] & 0.465 \\
        CM & 0.852 & 0.364 & 0.500 & 0.250 & 0.491 \\

        \midrule

        \multicolumn{6}{c}{MT-Text-CNN \cite{li2023automatic} (Baseline)}\\
        \midrule
        CC & 0.725 & 0.626 & 0.540 & 0.585 & 0.619\\
        IS & 0.486 & 0.457 & 0.432 & 0.437 & 0.453\\
        PS & 0.539 & 0.441 & 0.461 & 0.325 & 0.441\\
        CM & 0.536 & 0.659 & 0.449 & 0.255 & 0.475\\
        \midrule
        
        \multicolumn{6}{c}{BERT}\\
        \midrule
        CC & 0.885 & \cellcolor{gray!25}0.668 & \cellcolor{gray!25}0.644 & 0.426 & \cellcolor{gray!25}0.656\\
        IS & \cellcolor{gray!25}0.902 & \cellcolor{gray!25}0.766 & \cellcolor{gray!25}0.791 & 0.419 & \cellcolor{gray!25}0.719\\
        PS & \cellcolor{gray!25}0.842 & \cellcolor{gray!25}0.625 & \cellcolor{gray!25}0.727 & 0.000\footnotemark[\value{footnote}] & \cellcolor{gray!25}0.549\\
        CM & \cellcolor{gray!25}0.882 & 0.667 & 0.400 & 0.333 & 0.571\\

        \midrule
        
        \multicolumn{6}{c}{\textbf{BERT+AugGPT} (C/D not augmented)}\\
        \midrule
        CC & \cellcolor{gray!25}0.885 & \cellcolor{blue!25}0.925 & \cellcolor{blue!25}0.925 & \cellcolor{blue!25}0.796 & \cellcolor{blue!25}0.882\\
        IS & \cellcolor{gray!25}0.902 & \cellcolor{blue!25}0.922 & \cellcolor{blue!25}0.922 & \cellcolor{blue!25}0.851 & \cellcolor{blue!25}0.899\\
        PS & \cellcolor{gray!25}0.842 & \cellcolor{blue!25}0.895 & \cellcolor{blue!25}0.851 & \cellcolor{blue!25}0.842 & \cellcolor{blue!25}0.876\\
        CM & \cellcolor{gray!25}0.882 & \cellcolor{blue!25}0.826 & \cellcolor{blue!25}0.841 & \cellcolor{blue!25}0.840 & \cellcolor{blue!25}0.847\\
        \bottomrule
    \end{tabular}
    \label{tab:tb_bert_result}
\end{table}

\footnotetext{This result indicates that the model was unable to categorize the REQ because the amount of data is too small.}

\begin{somebox}
  \textbf{Answer to RQ2:}  \\
  Using BiLSTM+AugGPT for the \textit{identification} of SATD, and BERT+AugGPT for its \textit{categorization} mostly improve the F1-score of the analyzed baselines.
\end{somebox}


\subsection{RQ3: Keywords for SATD identification and categorization}
\label{answer-rq3}
Understanding how SATD keywords are used across different artifacts can reveal valuable insights into their similarities and differences. Keywords offer a lightweight method for identifying and categorizing SATD from various sources, aiding developers in comprehending technical debt (TD) \cite{chen2021multiclass}. For instance, keywords can generate reports that identify SATD types or highlight areas likely to contain SATD. 

There are different methods for extracting keywords. Potdar et al. \cite{potdar2014exploratory} manually summarized 62 common SATD keyword patterns for identifying SATD in source code comments. However, this manual approach is time-consuming and labor-intensive, making it challenging to encompass all possible SATD patterns. 

Recently, Li et al. \cite{li2023automatic} employed a backtracking approach to retrieve keywords by categorizing the most significant features using a trained Text-CNN. Their method requires a backtracking mechanism to map these features back into the architecture structure to discover keywords \cite{wang2020detecting}. However, backtracking can incur high computational costs due to its tendency to explore numerous partial solutions before arriving at a comprehensive solution \cite{russell2020ai4}.



Consequently, we employed a lightweight technique called KeyBERT \cite{grootendorst2020keybert} for keyword extraction and extraction-based summarization. KeyBERT utilizes BERT embeddings to identify keywords and keyphrases closely related to a given document. 
We used the original dataset provided by Li et al. \cite{li2023automatic} for keyword extraction, rather than the augmented dataset, to maintain data accuracy, avoid potential bias, and ensure the credibility of the extracted keywords. The search for SATD and Not-SATD examples in the artifacts is based on categorizing keywords strongly associated with the presence of SATD. Table~\ref{tab:tb_unigram_artifacts} summarizes the top SATD keywords extracted from the four artifacts. This table shows the most representative keywords identified by KeyBERT for various types of SATD in the artifacts. The complete list of keywords with similarity scores is available in the online repository \cite{satdaug}. 

\begin{table}[htpb!]
    \caption{Top SATD keywords from the artifacts} 
    \begin{tabular}{p{1.4cm} p{1.9cm} p{1.8cm} p{1.8cm}}
        \toprule
        CC & IS & PS & CM\\
        \midrule
        todo & \underline{patch} & test & typo\\
        fixme & test & nit & test\\
        \underline{method} & exception & code & fix typo\\
        class & checkstyle & unit test & fix\\
        \underline{exception} & fix & method & cleanup\\
        test & warning & line & remove\\
        \underline{null} & documentation & remove & polished\\
        hack & \underline{cluster} & comment & logging\\
        model & log & typo & code\\
        need & file & \underline{null} & \underline{patch}\\
        
        \bottomrule
    \end{tabular}
    \label{tab:tb_unigram_artifacts}
\end{table}

We noticed that certain keywords (\underline{underlined} in Table~\ref{tab:tb_unigram_artifacts}) were not uncovered by Li et al. \cite{li2023automatic}. These additional keywords seem to have been missed potentially due to the relatively low F1-score (0.611) achieved by their approach. Our approach can thus serve as a valuable supplement, automatically identifying SATD and highlighting keywords or phrases that aid in this identification. The top 10 keywords for each specific SATD type extracted using KeyBERT are listed in Table~\ref{tab:tb_unigram_types} and can be considered as patterns. Since C/D debt items outnumber other debt types, they are the most prevalent type of SATD across various sources. Consequently, the keywords associated with C/D debt predominantly overlap with the top keywords listed in Table~\ref{tab:tb_unigram_artifacts}.

\begin{table}[htp]
    \caption{Top keywords for specific types of SATD} 
    \begin{tabular}{p{1.2cm} p{1.8cm} p{1.7cm} p{2cm}}
        \toprule

        C/D & DOC & TES & REQ\\
        \midrule
        patch & typo & test & todo\\
        test & documentation & patch & fixme\\
        exception & patch & unit test & todo implement\\
        checkstyle & fix & flaky & todo check\\
        remove & doc & flaky test & implement\\
        warning & fix typo & coverage & implemented\\
        leak & comment & test case & check\\
        log & wiki & test coverage & null\\
        code & todo & testing & thread safe\\
        fix & tutorial & add test & thread\\
        \bottomrule
    \end{tabular}
    \label{tab:tb_unigram_types}
\end{table}

\begin{somebox}
  \textbf{Answer to RQ3:}  \\
  We assessed KeyBERT as capable of identifying keywords indicating SATD, including uncovering some specific keywords that were not extracted by Li et al.\cite{li2023automatic}. These keywords can be used to search for SATD manually through various artifacts, thus complementing automated detection approaches.
\end{somebox}

\section{Discussion}\label{sec5}
In the next subsections, we discuss the implications of our findings for both categories of stakeholders. 


\subsection{Implications for researchers and practitioners}


Utilizing deep learning approaches and data augmentation techniques like AugGPT for identifying and categorizing SATD has significant implications for developers and researchers. The effectiveness of identifying and categorizing SATD is influenced by both the data augmentation technique and the algorithms employed. This is evident when comparing methods such as XGBoost+SMOTE, XGBoost+EDA, JSD-GAN, MT-Text-CNN, and BERT. BERT excels in categorizing C/D and DOC debt, and TES debt (except in the CM artifact). Notably, the data augmentation technique has a greater impact, as seen in the superior results of BERT+AugGPT compared to BERT alone. Thus, alongside optimal deep learning architectures, prioritizing dataset quality is crucial for addressing data imbalances in SATD identification and categorization.

Improved identification and categorization of SATD instances, as demonstrated by our approach, enable developers to pinpoint problem areas in the code base and fix them, enhancing overall software quality. Although our approach outperforms baselines and state-of-practice methods, we recommend that software practitioners provide clear and standardized annotations in each artifact to ensure appropriate label quality using the right keywords (see results of RQ3).

In software development, `\textit{TODO}' and `\textit{fixme}' are keywords commonly used within source code to indicate tasks or issues that need to be addressed later. These keywords act as reminders for developers, highlighting areas of the code that require further attention or work. As presented in Table~\ref{tab:tb_unigram_artifacts}, our experimentation identifies these two keywords as the most prevalent within CC artifacts. Unfortunately, there is a lack of standardization or universally agreed-upon keywords for other artifacts. Consequently, we strongly advocate that practitioners consider using keywords in the IS, PS, and CM artifacts when introducing short-term solutions (i.e., SATD) into a project due to unavoidable circumstances. Developers are encouraged to utilize clear keywords based on each artifact to facilitate the subsequent categorization of SATD items.



Additionally, SATD identification datasets commonly exhibit class imbalance, where Not SATD items significantly outnumber SATD items. Addressing this imbalance is crucial for improving performance. 
We recommend that researchers contribute to developing and utilizing various techniques, such as transfer learning and few-shot learning. Furthermore, collaboration with software engineering experts and domain specialists is essential for contextual understanding of SATD and optimizing identification models. 

Finally, the approach shown in Subsection~\ref{sec:_data_augmentation} is general enough to be expanded outside the scope of SATD identification and categorization and utilized in other fields of Software Engineering research. Imbalanced datasets pose challenges for deep learning models, often resulting in strong performance for majority classes but weakness in minority classes. AugGPT, as demonstrated in our study, effectively augments minority class instances, diversifies the sample space, mitigates imbalance, and enhances overall model performance. 

\subsection{Threats to Validity}
\textit{Construct validity}. Threats to construct validity pertain to the extent to which operational measures accurately represent what is being investigated, following the research questions. In this study, we employ a widely used evaluation metric for imbalanced datasets, namely the F1-score, to evaluate the performance of our proposed method. We believe there is little threat to construct validity, as previous studies also use it to evaluate the performance \cite{da2017using, li2022identifying, li2023automatic, sharma2022self}.

\textit{Reliability}. The reliability threat may come from the datasets, namely the source code comments originally from Maldonado et al. \cite{da2017using} and the rest from Li et al. \cite{li2023automatic}. It is worth noting that the datasets were labeled manually, introducing the possibility of personal biases. However, it is encouraging to find that the results of manual classification performed by different individuals, as reported in the prior studies, demonstrate a high degree of consistency. The calculated Cohen's Kappa coefficients, which reached +0.81 and +0.74, further support this consistency and indicate that the data sets possess a reasonable level of reliability that can help mitigate concerns about potential inconsistencies caused by manual classification.

\textit{External validity}. While our experiments used publicly available datasets, the generalizability of our approach to other projects and programming languages remains uncertain, which may limit its broader applicability. However, we used diverse datasets from various open-source projects, covering different contributors, SLOC, comments, and SATD types. Since our method processes natural language text from code comments, issue trackers, pull requests, and commit messages, these artifacts remain relevant across different languages.

Further research is needed to validate our approach with a larger, more diverse dataset. Moreover, as our training data comes from open-source projects, there are limitations in generalizing results to industry projects, which may handle technical debt differently. While our findings may extend to similar open-source projects, SATD documentation practices in industry could vary.

\section{Conclusion and Future Work}\label{sec6}
This study aims to automatically identify and categorize SATD in software artifacts, such as source code comments, issues section, pull requests, and commit messages. Our approach employs BiLSTM for initial identification and BERT for subsequent categorization. Analysis of the datasets revealed a significant imbalance in SATD types, addressed through AugGPT for data augmentation, which markedly improves F1-score performance over baselines. Our method provides a robust framework for comprehensive SATD identification and categorization across varied sources and also offers insights into key indicative keywords for each artifact and SATD type.

There are several potential research topics for future research. The current study addresses class imbalance using a large language model (LLM)-based data augmentation strategy. Future research could explore alternative methods of dealing with imbalanced data in SATD detection, such as few-shot learning or transfer learning to further improve detection accuracy, particularly for rare SATD categories. Additionally, we encourage investigating the use of other LLMs, such as Gemma-7B, Vicuna-13B, Falcon 180B, and LLaMa 3-8B, to address data imbalance challenges. Finally, investigating keywords and patterns from several software artifacts to enhance understanding and explainability of SATD represents an interesting research direction.


\section*{Acknowledgement}
This work was financially supported by the Indonesian Education Scholarship (BPI) from the Center for Financing of Higher Education (BPPT) and Indonesia Endowment Fund for Education (LPDP).



\balance
\bibliography{main}

\begin{thebibliography}{10}
\providecommand{\url}[1]{#1}
\csname url@samestyle\endcsname
\providecommand{\newblock}{\relax}
\providecommand{\bibinfo}[2]{#2}
\providecommand{\BIBentrySTDinterwordspacing}{\spaceskip=0pt\relax}
\providecommand{\BIBentryALTinterwordstretchfactor}{4}
\providecommand{\BIBentryALTinterwordspacing}{\spaceskip=\fontdimen2\font plus
\BIBentryALTinterwordstretchfactor\fontdimen3\font minus \fontdimen4\font\relax}
\providecommand{\BIBforeignlanguage}[2]{{%
\expandafter\ifx\csname l@#1\endcsname\relax
\typeout{** WARNING: IEEEtran.bst: No hyphenation pattern has been}%
\typeout{** loaded for the language `#1'. Using the pattern for}%
\typeout{** the default language instead.}%
\else
\language=\csname l@#1\endcsname
\fi
#2}}
\providecommand{\BIBdecl}{\relax}
\BIBdecl

\bibitem{cunningham1992wycash}
W.~Cunningham, ``The wycash portfolio management system,'' \emph{ACM SIGPLAN OOPS Messenger}, vol.~4, no.~2, pp. 29--30, 1992.

\bibitem{avgeriou2016managing}
P.~Avgeriou, P.~Kruchten, I.~Ozkaya, and C.~Seaman, ``Managing technical debt in software engineering (dagstuhl seminar 16162),'' in \emph{Dagstuhl reports}, vol.~6, no.~4.\hskip 1em plus 0.5em minus 0.4em\relax Deutsch: Schloss Dagstuhl-Leibniz-Zentrum fuer Informatik, 2016.

\bibitem{brown2010managing}
N.~Brown, Y.~Cai, Y.~Guo, R.~Kazman, M.~Kim, P.~Kruchten, E.~Lim, A.~MacCormack, R.~Nord, I.~Ozkaya \emph{et~al.}, ``Managing technical debt in software-reliant systems,'' in \emph{Proceeding of the FSE/SDP workshop on Future of software engineering research}.\hskip 1em plus 0.5em minus 0.4em\relax Santa Fe New Mexico, USA: ACM, 2010, pp. 47--52.

\bibitem{marinescu2004detection}
R.~Marinescu, ``Detection strategies: Metrics-based rules for detecting design flaws,'' in \emph{20th IEEE Int. Conf. on Software Maintenance, 2004. Proceedings.}\hskip 1em plus 0.5em minus 0.4em\relax IEEE, 2004, pp. 350--359.

\bibitem{marinescu2010incode}
R.~Marinescu, G.~Ganea, and I.~Verebi, ``Incode: Continuous quality assessment and improvement,'' in \emph{2010 14th European Conf. on Software Maintenance and Reengineering}.\hskip 1em plus 0.5em minus 0.4em\relax IEEE, 2010, pp. 274--275.

\bibitem{zazworka2014comparing}
N.~Zazworka, A.~Vetro’, C.~Izurieta, S.~Wong, Y.~Cai, C.~Seaman, and F.~Shull, ``Comparing four approaches for technical debt identification,'' \emph{Software Quality J}, vol.~22, pp. 403--426, 2014.

\bibitem{potdar2014exploratory}
A.~Potdar and E.~Shihab, ``An exploratory study on self-admitted technical debt,'' in \emph{2014 IEEE ICSME}.\hskip 1em plus 0.5em minus 0.4em\relax IEEE, 2014, pp. 91--100.

\bibitem{huang2018identifying}
Q.~Huang, E.~Shihab, X.~Xia, D.~Lo, and S.~Li, ``Identifying self-admitted technical debt in open source projects using text mining,'' \emph{Empir. Softw. Eng.}, vol.~23, pp. 418--451, 2018.

\bibitem{fontana2016antipattern}
F.~A. Fontana, J.~Dietrich, B.~Walter, A.~Yamashita, and M.~Zanoni, ``Antipattern and code smell false positives: Preliminary conceptualization and classification,'' in \emph{2016 IEEE 23rd SANER}, vol.~1.\hskip 1em plus 0.5em minus 0.4em\relax IEEE, 2016, pp. 609--613.

\bibitem{da2017using}
E.~da~Silva~Maldonado, E.~Shihab, and N.~Tsantalis, ``Using natural language processing to automatically detect self-admitted technical debt,'' \emph{IEEE TSE}, vol.~43, no.~11, pp. 1044--1062, 2017.

\bibitem{dai2017detecting}
K.~Dai and P.~Kruchten, ``Detecting technical debt through issue trackers.'' in \emph{QuASoQ@ APSEC}, 2017, pp. 59--65.

\bibitem{wattanakriengkrai2019automatic}
S.~Wattanakriengkrai, N.~Srisermphoak, S.~Sintoplertchaikul, M.~Choetkiertikul, C.~Ragkhitwetsagul, T.~Sunetnanta, H.~Hata, and K.~Matsumoto, ``Automatic classifying self-admitted technical debt using n-gram idf,'' in \emph{2019 26th APSEC}.\hskip 1em plus 0.5em minus 0.4em\relax IEEE, 2019, pp. 316--322.

\bibitem{xavier2020beyond}
L.~Xavier, F.~Ferreira, R.~Brito, and M.~T. Valente, ``Beyond the code: Mining self-admitted technical debt in issue tracker systems,'' in \emph{Proceeding of the 17th Int. Conf. on MSR}, 2020, pp. 137--146.

\bibitem{zhu2023scgru}
K.~Zhu, M.~Yin, D.~Zhu, X.~Zhang, C.~Gao, and J.~Jiang, ``Scgru: A general approach for identifying multiple classes of self-admitted technical debt with text generation oversampling,'' \emph{JSS}, vol. 195, p. 111514, 2023.

\bibitem{alves2014towards}
N.~S. Alves, L.~F. Ribeiro, V.~Caires, T.~S. Mendes, and R.~O. Spínola, ``Towards an ontology of terms on technical debt,'' in \emph{2014 Sixth Int. Workshop on MTD}.\hskip 1em plus 0.5em minus 0.4em\relax Victoria, BC, Canada: IEEE, 2014, pp. 1--7.

\bibitem{maldonado2015detecting}
E.~d.~S. Maldonado and E.~Shihab, ``Detecting and quantifying different types of self-admitted technical debt,'' in \emph{2015 IEEE 7Th Int. workshop on MTD}.\hskip 1em plus 0.5em minus 0.4em\relax IEEE, 2015, pp. 9--15.

\bibitem{chen2021multiclass}
X.~Chen, D.~Yu, X.~Fan, L.~Wang, and J.~Chen, ``Multiclass classification for self-admitted technical debt based on xgboost,'' \emph{IEEE Transactions on Reliability}, vol.~71, no.~3, pp. 1309--1324, 2021.

\bibitem{li2022identifying}
Y.~Li, M.~Soliman, and P.~Avgeriou, ``Identifying self-admitted technical debt in issue tracking systems using machine learning,'' \emph{Empir. Softw. Eng.}, vol.~27, no.~6, p. 131, 2022.

\bibitem{xavier2022documentation}
L.~Xavier, J.~E. Montandon, F.~Ferreira, R.~Brito, and M.~T. Valente, ``On the documentation of self-admitted technical debt in issues,'' \emph{Empir. Softw. Eng.}, vol.~27, no.~7, p. 163, 2022.

\bibitem{li2023automatic}
Y.~Li, M.~Soliman, and P.~Avgeriou, ``Automatic identification of self-admitted technical debt from four different sources,'' \emph{Empir. Softw. Eng.}, vol.~28, no.~3, pp. 1--38, 2023.

\bibitem{provost2013data}
F.~Provost and T.~Fawcett, \emph{Data Science for Business: What you need to know about data mining and data-analytic thinking}.\hskip 1em plus 0.5em minus 0.4em\relax " O'Reilly Media, Inc.", 2013.

\bibitem{satdaug}
E.~Sutoyo, P.~Avgeriou, and A.~Capiluppi, ``Replication package of deep learning and data augmentation for detecting self-admitted technical debt,'' \url{https://github.com/edisutoyo/satd-augmentation}, 2024.

\bibitem{lim2012balancing}
E.~Lim, N.~Taksande, and C.~Seaman, ``A balancing act: What software practitioners have to say about technical debt,'' \emph{IEEE software}, vol.~29, no.~6, pp. 22--27, 2012.

\bibitem{wehaibi2016examining}
S.~Wehaibi, E.~Shihab, and L.~Guerrouj, ``Examining the impact of self-admitted technical debt on software quality,'' in \emph{2016 IEEE 23Rd SANER}, vol.~1.\hskip 1em plus 0.5em minus 0.4em\relax IEEE, 2016, pp. 179--188.

\bibitem{li2015systematic}
Z.~Li, P.~Avgeriou, and P.~Liang, ``A systematic mapping study on technical debt and its management,'' \emph{JSS}, vol. 101, pp. 193--220, 2015.

\bibitem{sala2021debthunter}
I.~Sala, A.~Tommasel, and F.~Arcelli~Fontana, ``Debthunter: A machine learning-based approach for detecting self-admitted technical debt,'' in \emph{Evaluation and Assessment in Software Engineering}, 2021, pp. 278--283.

\bibitem{guo2021far}
Z.~Guo, S.~Liu, J.~Liu, Y.~Li, L.~Chen, H.~Lu, and Y.~Zhou, ``How far have we progressed in identifying self-admitted technical debts? a comprehensive empirical study,'' \emph{ACM TOSEM}, vol.~30, no.~4, pp. 1--56, 2021.

\bibitem{sridharan2021data}
M.~Sridharan, M.~Mantyla, L.~Rantala, and M.~Claes, ``Data balancing improves self-admitted technical debt detection,'' in \emph{2021 IEEE/ACM 18th Int. Conf. on MSR}.\hskip 1em plus 0.5em minus 0.4em\relax IEEE, 2021, pp. 358--368.

\bibitem{zhu2021detecting}
K.~Zhu, M.~Yin, and Y.~Li, ``Detecting and classifying self-admitted of technical debt with cnn-bilstm,'' in \emph{J. of Physics: Conference Series}, vol. 1955, no.~1.\hskip 1em plus 0.5em minus 0.4em\relax IOP Publishing, 2021, p. 012102.

\bibitem{yu2022exploiting}
J.~Yu, K.~Zhao, J.~Liu, X.~Liu, Z.~Xu, and X.~Wang, ``Exploiting gated graph neural network for detecting and explaining self-admitted technical debts,'' \emph{JSS}, vol. 187, p. 111219, 2022.

\bibitem{zampetti2021self}
F.~Zampetti, G.~Fucci, A.~Serebrenik, and M.~Di~Penta, ``Self-admitted technical debt practices: a comparison between industry and open-source,'' \emph{Empir. Softw. Eng.}, vol.~26, pp. 1--32, 2021.

\bibitem{sutoyo2023detecting}
E.~Sutoyo and A.~Capiluppi, ``Self-admitted technical debt detection approaches: A decade systematic review,'' \emph{arXiv preprint arXiv:2312.15020}, 2024.

\bibitem{drummond2003c4}
C.~Drummond, R.~C. Holte \emph{et~al.}, ``C4. 5, class imbalance, and cost sensitivity: why under-sampling beats over-sampling,'' in \emph{Workshop on learning from imbalanced datasets II}, vol.~11, 2003, pp. 1--8.

\bibitem{yap2014application}
B.~W. Yap, K.~A. Rani, H.~A.~A. Rahman, S.~Fong, Z.~Khairudin, and N.~N. Abdullah, ``An application of oversampling, undersampling, bagging and boosting in handling imbalanced datasets,'' in \emph{Proceeding of the first Int. Conf. on DaEng-2013}.\hskip 1em plus 0.5em minus 0.4em\relax Springer, 2014, pp. 13--22.

\bibitem{lee2018oversampling}
S.-K. Lee, S.-J. Hong, and S.-I. Yang, ``Oversampling for imbalanced data classification using adversarial network,'' in \emph{2018 Int. Conf. on ICTC}.\hskip 1em plus 0.5em minus 0.4em\relax IEEE, 2018, pp. 1255--1257.

\bibitem{feng2021survey}
S.~Y. Feng, V.~Gangal, J.~Wei, S.~Chandar, S.~Vosoughi, T.~Mitamura, and E.~Hovy, ``A survey of data augmentation approaches for nlp,'' \emph{arXiv preprint arXiv:2105.03075}, 2021.

\bibitem{sennrich2015improving}
R.~Sennrich, B.~Haddow, and A.~Birch, ``Improving neural machine translation models with monolingual data,'' \emph{arXiv preprint arXiv:1511.06709}, 2015.

\bibitem{jindal2020augmenting}
A.~Jindal, A.~G. Chowdhury, A.~Didolkar, D.~Jin, R.~Sawhney, and R.~Shah, ``Augmenting nlp models using latent feature interpolations,'' in \emph{Proceedings of the 28th Int. Conf. on Computational Linguistics}, 2020, pp. 6931--6936.

\bibitem{dai2023chataug}
H.~Dai, Z.~Liu, W.~Liao, X.~Huang, Z.~Wu, L.~Zhao, W.~Liu, N.~Liu, S.~Li, D.~Zhu \emph{et~al.}, ``Auggpt: Leveraging chatgpt for text data augmentation,'' \emph{arXiv preprint arXiv:2302.13007}, 2023.

\bibitem{brown2020language}
T.~Brown, B.~Mann, N.~Ryder, M.~Subbiah, J.~D. Kaplan, P.~Dhariwal, A.~Neelakantan, P.~Shyam, G.~Sastry, A.~Askell \emph{et~al.}, ``Language models are few-shot learners,'' \emph{Advances in neural information processing systems}, vol.~33, pp. 1877--1901, 2020.

\bibitem{white2023prompt}
J.~White, Q.~Fu, S.~Hays, M.~Sandborn, C.~Olea, H.~Gilbert, A.~Elnashar, J.~Spencer-Smith, and D.~C. Schmidt, ``A prompt pattern catalog to enhance prompt engineering with chatgpt,'' \emph{arXiv preprint arXiv:2302.11382}, 2023.

\bibitem{shannon2001mathematical}
C.~E. Shannon, ``A mathematical theory of communication,'' \emph{ACM SIGMOBILE mobile computing and communications review}, vol.~5, no.~1, pp. 3--55, 2001.

\bibitem{de2006documentation}
S.~C.~B. de~Souza, N.~Anquetil, and K.~M. de~Oliveira, ``Which documentation for software maintenance?'' \emph{J. of the Brazilian Computer Society}, vol.~12, pp. 31--44, 2006.

\bibitem{chowdhary2020natural}
K.~Chowdhary and K.~Chowdhary, ``Natural language processing,'' \emph{Fundamentals of artificial intelligence}, pp. 603--649, 2020.

\bibitem{pennington2014glove}
J.~Pennington, R.~Socher, and C.~D. Manning, ``Glove: Global vectors for word representation,'' in \emph{Proceeding of the 2014 Conf. on EMNLP}, 2014, pp. 1532--1543.

\bibitem{devlin2018bert}
J.~Devlin, M.-W. Chang, K.~Lee, and K.~Toutanova, ``Bert: Pre-training of deep bidirectional transformers for language understanding,'' \emph{arXiv preprint arXiv:1810.04805}, 2018.

\bibitem{schuster1997bidirectional}
M.~Schuster and K.~K. Paliwal, ``Bidirectional recurrent neural networks,'' \emph{IEEE transactions on Signal Processing}, vol.~45, no.~11, pp. 2673--2681, 1997.

\bibitem{loshchilov2017decoupled}
I.~Loshchilov and F.~Hutter, ``Decoupled weight decay regularization,'' \emph{arXiv preprint arXiv:1711.05101}, 2017.

\bibitem{yin2023two}
M.~Yin, J.~Wang, D.~Zhu, and C.~Gao, ``A two-stage approach for identifying and interpreting self-admitted technical debt,'' \emph{Applied Intelligence}, vol.~53, no.~22, pp. 26\,592--26\,602, 2023.

\bibitem{yu2023detecting}
J.~Yu, X.~Zhou, X.~Liu, J.~Liu, Z.~Xie, and K.~Zhao, ``Detecting multi-type self-admitted technical debt with generative adversarial network-based neural networks,'' \emph{IST}, vol. 158, p. 107190, 2023.

\bibitem{wang2020detecting}
X.~Wang, J.~Liu, L.~Li, X.~Chen, X.~Liu, and H.~Wu, ``Detecting and explaining self-admitted technical debts with attention-based neural networks,'' in \emph{Proceeding of the 35th IEEE/ACM Int. Conf. on Autom. Softw. Eng.}, 2020, pp. 871--882.

\bibitem{russell2020ai4}
S.~Russell and P.~Norvig, \emph{Artificial Intelligence: A Modern Approach}, 4th~ed.\hskip 1em plus 0.5em minus 0.4em\relax Pearson Education Limited, 2020.

\bibitem{grootendorst2020keybert}
\BIBentryALTinterwordspacing
M.~Grootendorst, ``Keybert: Minimal keyword extraction with bert.'' 2020. [Online]. Available: \url{https://doi.org/10.5281/zenodo.4461265}
\BIBentrySTDinterwordspacing

\bibitem{sharma2022self}
R.~Sharma, R.~Shahbazi, F.~H. Fard, Z.~Codabux, and M.~Vidoni, ``Self-admitted technical debt in r: detection and causes,'' \emph{Autom. Softw. Eng.}, vol.~29, no.~2, p.~53, 2022.

\end{thebibliography}
\bibliographystyle{IEEEtran} 

\end{document}